\documentclass[preprint]{aastex63}
\usepackage{graphicx}
\begin{document}

\title{Absorption Line Observations of H$_3^+$ and CO in Sight Lines Toward the Vela and W28 Supernova Remnants}

\author[0000-0001-8533-6440]{Nick Indriolo}
\affil{AURA for the European Space Agency (ESA), ESA Office, Space Telescope Science Institute, 3700 San Martin Drive, Baltimore, MD 21218, USA}

\begin{abstract}
Supernova remnants act as particle accelerators, providing the cosmic-ray protons that permeate the interstellar medium and initiate the ion-molecule reactions that drive interstellar chemistry. Enhanced fluxes of cosmic-ray protons in close proximity to supernova remnants have been inferred from observations tracing particle interactions with nearby molecular gas. Here I present observations of H$_3^+$ and CO absorption, molecules that serve as tracers of the cosmic-ray ionization rate and gas density, respectively, in sight lines toward the W28 and Vela supernova remnants. Cosmic-ray ionization rates inferred from these observations range from about 2--10 times the average value in Galactic diffuse clouds ($\sim 3\times10^{-16}$ s$^{-1}$), suggesting that the gas being probed is experiencing an elevated particle flux. While it is difficult to constrain the line-of-sight location of the absorbing gas with respect to the supernova remnants, these results are consistent with a scenario where cosmic rays are diffusing away from the acceleration site and producing enhanced ionization rates in the surrounding medium.

\end{abstract}

\section{Introduction} \label{sec_intro}

The acceleration of cosmic rays occurs in magnetized shocks as particles scatter back and forth across the shock front \citep[][and references therein]{drury1983}. Various observations provide evidence for enhanced cosmic-ray fluxes in the immediate vicinities of supernova remnants (SNRs), thus confirming the long-held theory that such objects act as particle accelerators \citep{blasi2013}. These observations include gamma-ray emission co-incident with molecular material in close proximity to SNRs \citep[e.g.,][]{abdo2009w51,acciari2009,ackermann2013}---the result of $\pi^0$ production from inelastic scattering of hadronic cosmic-rays off of ambient nuclei---and enhanced abundances of particular molecular ions like H$_3^+$ and HCO$^+$ \citep[e.g.,][]{indriolo2010ic443,ceccarelli2011} that have formation rates directly linked to the cosmic-ray ionization rate.  Each of these observables traces slightly different energy regimes in the spectrum of cosmic-ray protons. Pion production ``turns on'' for protons with kinetic energy above about 280~MeV, and dominates the energy loss function for $E\gtrsim 1$~GeV \citep[][and references therein]{padovani2018}. Ionization of H and H$_2$, however, is dominated by particles with $E\sim 10$--100~MeV \citep{padovani2009}. The diffusion of cosmic rays away from the acceleration site is energy dependent, with higher energy particles escaping sooner and thus traveling farther than lower energy particles \citep{gabici2009}. As a result, observations of gamma-ray emission and molecular ion abundances can provide complementary information about the extent to which different energy particles have diffused away from an SNR.

Gamma-ray emission has been found to be spatially coincident with several SNRs \citep[e.g.,][]{acero2016}. Analyses of the gamma-ray spectra of these sources and of the spatial distribution of emission with respect to molecular and atomic gas have demonstrated that in many cases the gamma-ray emission is mostly hadronic in origin; i.e., from $\pi^0$ creation and decay following inelastic scattering of protons off of ambient nuclei. Among these objects are the W28 SNR \citep{aharonian2008,abdo2010w28}, RX J0852.0$-$4622 \citep{fukui2017}, and Puppis A \citep{aruga2022}. The latter two sources are coincident on sky with the Vela SNR, which hosts the Vela pulsar and pulsar wind nebula, both strong gamma-ray emitters \citep{abdo2009Velapulsar,abdo2010Velapwn,grondin2013}. While there is no gamma-ray emission associated with molecular clouds in the Vela SNR, the remnant does show hints of interaction between shocks and the ambient interstellar medium. Here, I present observations of H$_3^+$ and CO absorption in sight lines toward these SNRs. CO observations are used to place constraints on the properties of the absorbing clouds, and H$_3^+$ observations are used to infer cosmic-ray ionization rates.

\subsection{Vela SNR}

The Vela pulsar is located at a distance of 287$^{+19}_{-17}$~pc \citep{dodson2003}, and is within the Vela SNR. Although molecular gas is observed to be coincident with the Vela SNR on-sky \citep[e.g.,][]{moriguchi2001}, it is unclear how much of this gas is in close proximity to the SNR along the line of sight. Distance estimates place the Vela Molecular Ridge (VMR)---a complex of giant molecular clouds---at 700$\pm$200~pc \citep{murphy1991,liseau1992,yamaguchi1999}, well behind the Vela SNR, but there are also some indications of interactions between the SNR and small pockets of molecular gas \citep{nichols2004,miceli2005}. 

\begin{figure}
\epsscale{1.0}
\plotone{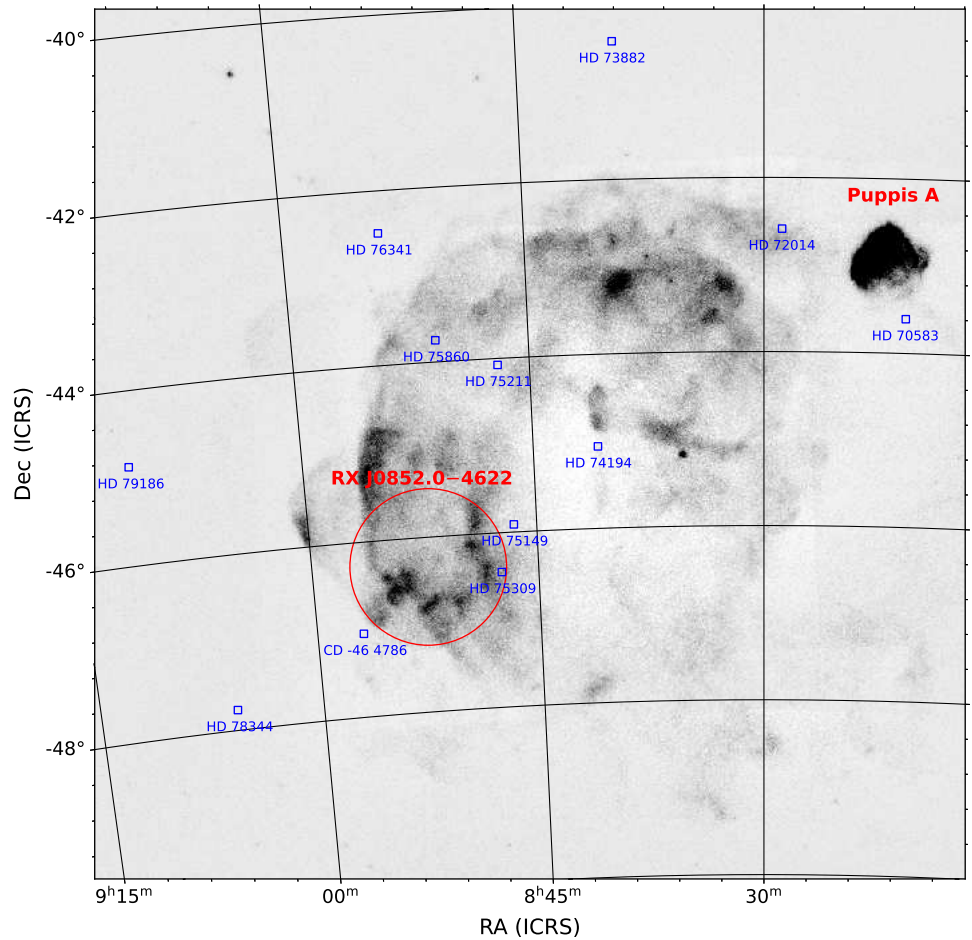}
\caption{ROSAT X-ray map of the Vela SNR for photon energies in the range $0.1\leq E\leq 2.5$~keV \citep{aschenbach1998}. Data were retrieved from the NASA High Energy Astrophysics Science Archive Research Center (https://heasarc.gsfc.nasa.gov/). Flux scaling is arbitrary as the intent of the image is simply to show the locations of target sight lines with respect to the supernova remnants. Sight lines targeted in this work, and previously observed sight lines with cosmic-ray ionization rate estimates, are marked by blue squares and are labeled. The location and size of RX J0852.0$-$4622 is marked by a red circle \citep{maxted2018}.}
\label{fig_vela_map}
\end{figure}

Two more SNRs coincident with the Vela SNR on-sky but at larger distances are of interest to this study: RX J0852.0$-$4622 and Puppis A. RX J0852.0$-$4622 and Puppis A are at distances of 700--800~pc \citep{allen2015} and 1.3$\pm$0.3~kpc \citep{reynoso2017}, respectively, and as mentioned above, both show gamma-ray emission that is indicative of cosmic-ray interactions with ambient gas \citep{aharonian2005,tanaka2011,fukui2017,maxted2018,hewitt2012,aruga2022}. Figure \ref{fig_vela_map} shows the Vela SNR region in X-ray emission \citep{aschenbach1998} and marks the locations of the sight lines targeted in this study and retrieved from the literature. Depending on the location of the absorbing gas along the target sight lines, it may be probing the cosmic-ray flux in close proximity to any of these three supernova remnants.

\subsection{W28 SNR}

\begin{figure}
\epsscale{1.0}
\plotone{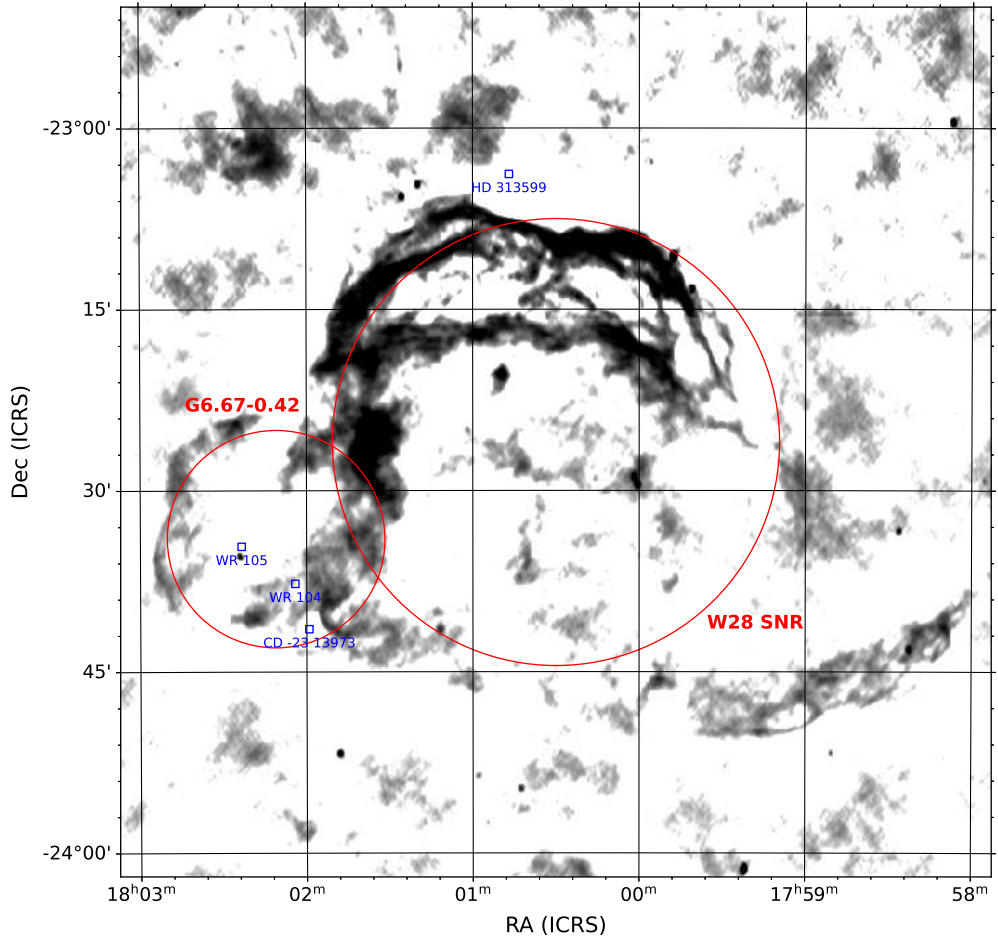}
\caption{VLA 90~cm continuum emission image of the W28 SNR. Data were retrieved from the MAGPIS website \citep[][http://third.ucllnl.org/gps]{helfand2006}. Flux scaling is arbitrary as the intent of the image is simply to show the locations of target sight lines with respect to the supernova remnants. Sight lines targeted in this work, and previously observed sight lines with cosmic-ray ionization rate estimates, are marked by blue squares and are labeled. The approximate sizes and locations of G6.67$-$0.42 and the W28 SNR are marked by red circles \citep{brogan2006}.}
\label{fig_w28_map}
\end{figure}

Distance estimates to the W28 SNR have varied from about 1.2 to 3.3~kpc \citep[][and references therein]{nicholas2012}. Observations targeting Na~\textsc{i} and Ca~\textsc{ii} absorption toward stars that lie within the SNR 90~cm continuum radio shell on-sky show high velocity components toward a star at 2.4$\pm$0.3~kpc, but no such components toward stars at about 1.7~kpc \citep{ritchey2020}. This is in good agreement with the kinematic distance of 1.9$\pm$0.3~kpc determined from H~\textsc{i} self absorption at $v_{\rm LSR}=7$~km~s$^{-1}$ \citep{velazquez2002}. The W28 SNR is known to be interacting with part of a molecular cloud on its northeast side \citep[][and references therein]{maxted2016}. The bulk of CO $J=$1--0 and 2--1 emission from this cloud is in the velocity range 0~km~s$^{-1}\leq v_{\rm LSR}\leq 20$~km~s$^{-1}$ \citep{fukui2008}, and the cloud is coincident with gamma-ray emission observed by HESS \citep{aharonian2008} and {\it Fermi}-LAT \citep{abdo2010w28}, indicating a large flux of cosmic rays entering the cloud from the nearby SNR. Cosmic-ray ionization rates inferred from observations of HCO$^+$ and DCO$^+$ emission in this cloud are of order a few times 10$^{-15}$~s$^{-1}$ \citep{vaupre2014}, about 100 times larger than the canonical value in dense clouds. The eastern edge of the W28 SNR partially overlaps with another remnant, G6.67$-$0.42 \citep{brogan2006}, estimated to be at a distance of 3.7~kpc \citep{wang2020}. Figure \ref{fig_w28_map} shows the W28 region in 90~cm continuum emission \citep{brogan2006} and marks the locations of sight lines targeted in this study. For a more detailed image of the region, including CO emission and gamma-ray emission, the reader is directed to Figure 1 of \citet{maxted2016}.

\section{Observations and Data Reduction}
All H$_3^+$ and CO observations presented herein were made using the Cryogenic Infrared Echelle Spectrograph \citep[CRIRES;][]{kaufl2004} at the Very Large Telescope (VLT). These observations utilized the 0\farcs2 slit to provide a resolving power (resolution) of about 100,000 (3~km~s$^{-1}$). For H$_3^+$ observations a reference wavelength of 3715~nm was used to place the $R(1,1)^u$ and $R(1,0)$ transitions on detector 1 and the $R(1,1)^l$ transition on detector 3. For CO observations a reference wavelength of 4681~nm was used to cover the $R(6)$ through $P(6)$ transitions of the $v=1-0$ band (note that the $P(4)$ and $R(3)$ transitions fall in gaps in wavelength coverage between the detectors). To maximize starlight passing through the narrow slit, the adaptive optics system was employed. Spectra were obtained in an ABBA nodding pattern, with 10\arcsec\ separation between nod positions, and 3\arcsec\ jitter about those positions. Observations of each science target were immediately preceded or followed by the observations of a bright, early type star at similar airmass for use as a telluric standard star.

Raw CRIRES images were processed using the CRIRES pipeline version 2.3.3. Standard calibration techniques, including subtraction of dark frames, division by flat fields, interpolation over bad pixels, and correction for detector non-linearity effects, were applied. Consecutive A and B nod position images were subtracted from each other to remove sky emission features, and all images from each nod position were combined to create average A and B images. Spectra were extracted from these images using the \texttt{apall} routine in \textsc{iraf}\footnote{https://iraf-community.github.io/} and then imported to IGOR Pro.\footnote{https://www.wavemetrics.com} Wavelength calibration was performed using atmospheric absorption lines, and is accurate to $\pm1$~km~s$^{-1}$. Spectra from the A and B nod positions were then averaged onto a common wavelength scale.

\begin{figure}
\epsscale{1.0}
\plotone{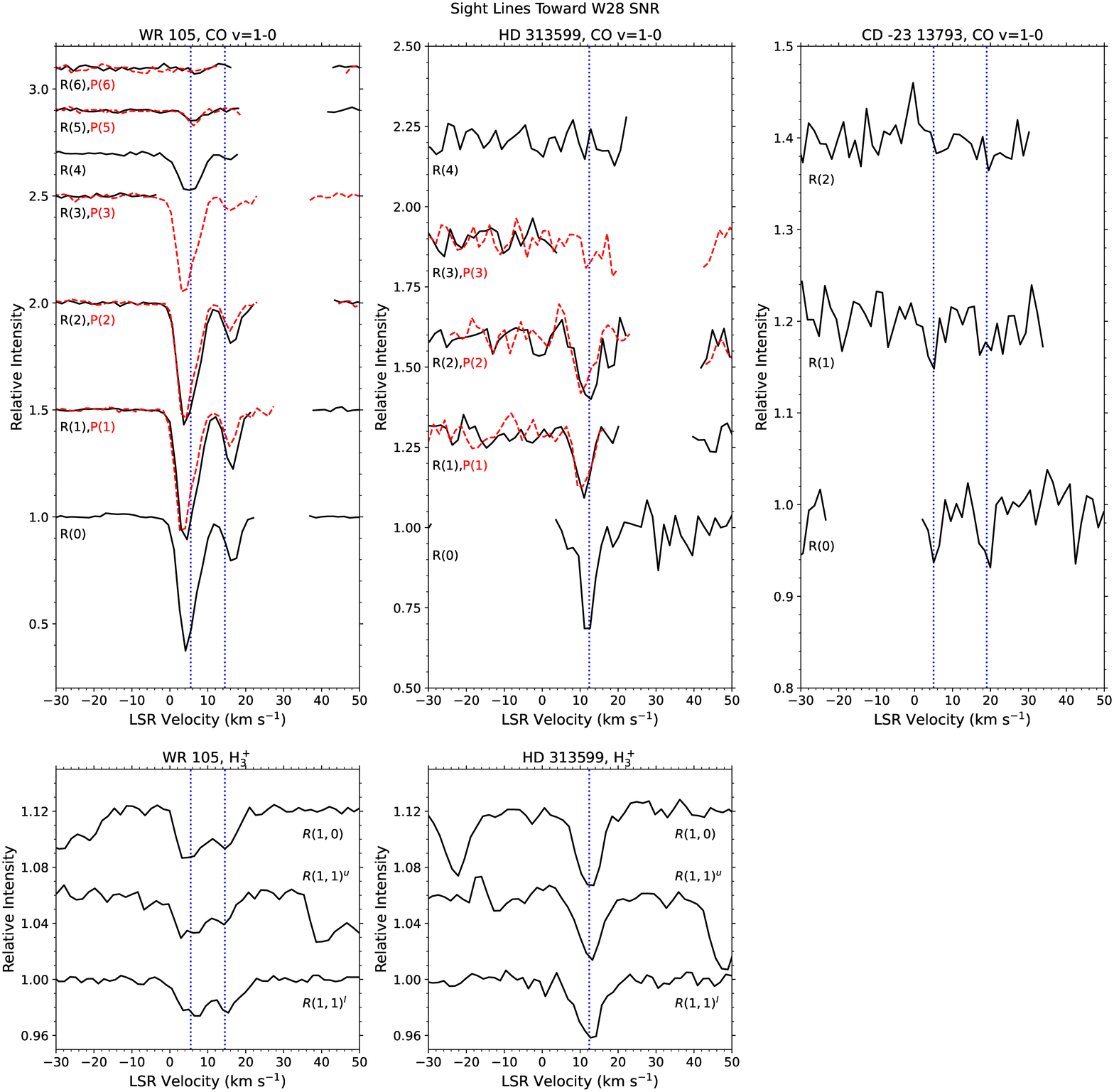}
\caption{The top panels show spectra of three sight lines toward the W28 SNR---WR 105, HD 313599, and CD $-$23 13793---targeting the $v=1$--0 band of CO near 4.7~$\mu$m. $R$-branch lines are shown as solid black curves, and $P$-branch lines as dashed red curves. The bottom panels show spectra in the same sight lines (except CD $-$23 13793, which had low S/N) targeting the $\nu_2$ band of H$_3^+$. In all panels spectra have been shifted vertically for clarity, and gaps indicate regions where the removal of atmospheric features was poor.}
\label{fig_w28_spectra}
\end{figure}

\begin{figure}
\epsscale{1.0}
\plotone{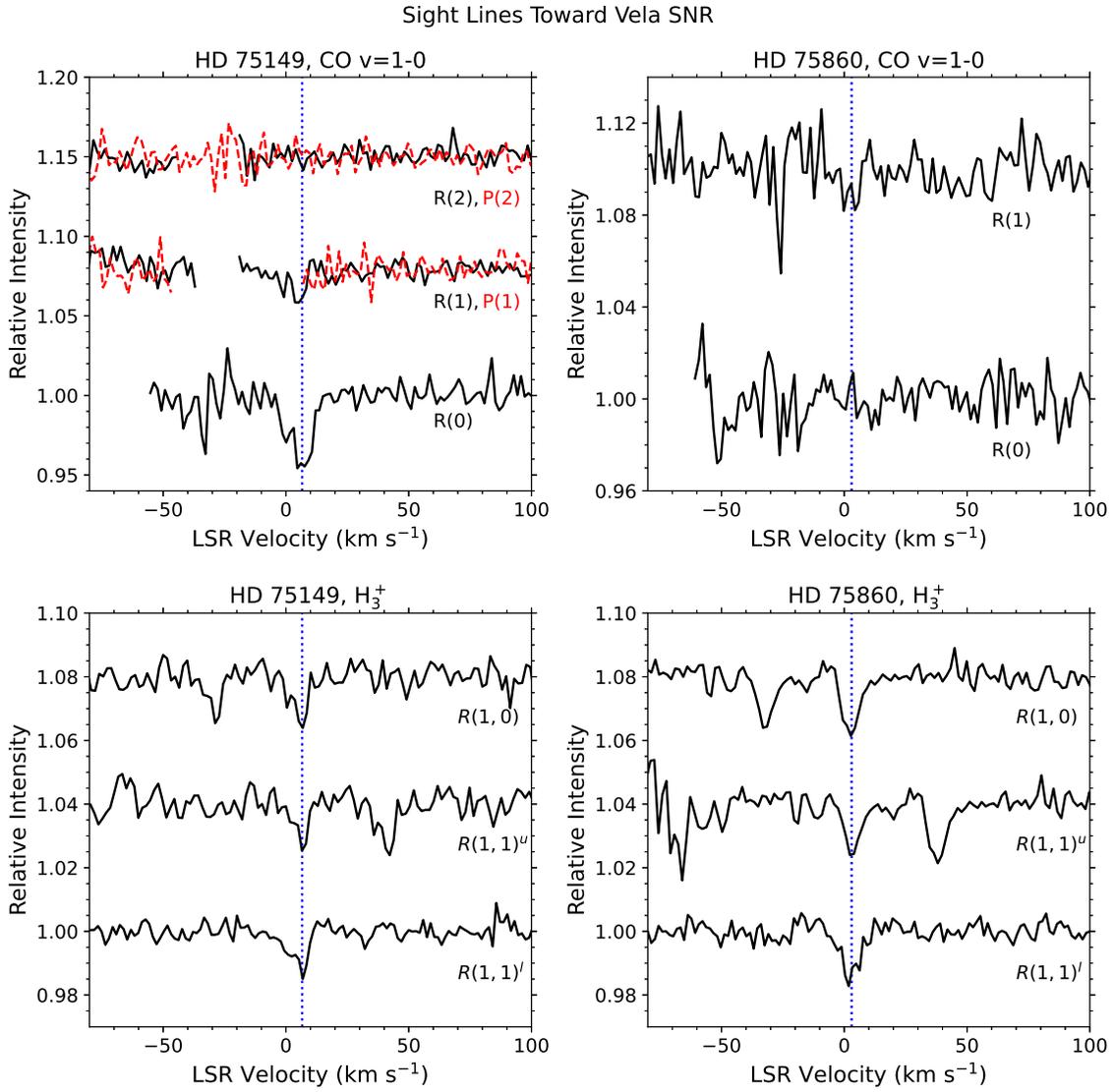}
\caption{Same as Figure \ref{fig_w28_spectra}, except for two sight lines---HD 75149 and HD 75860---toward the Vela SNR.}
\label{fig_vela_spectra}
\end{figure}

\begin{figure}
\epsscale{1.0}
\plotone{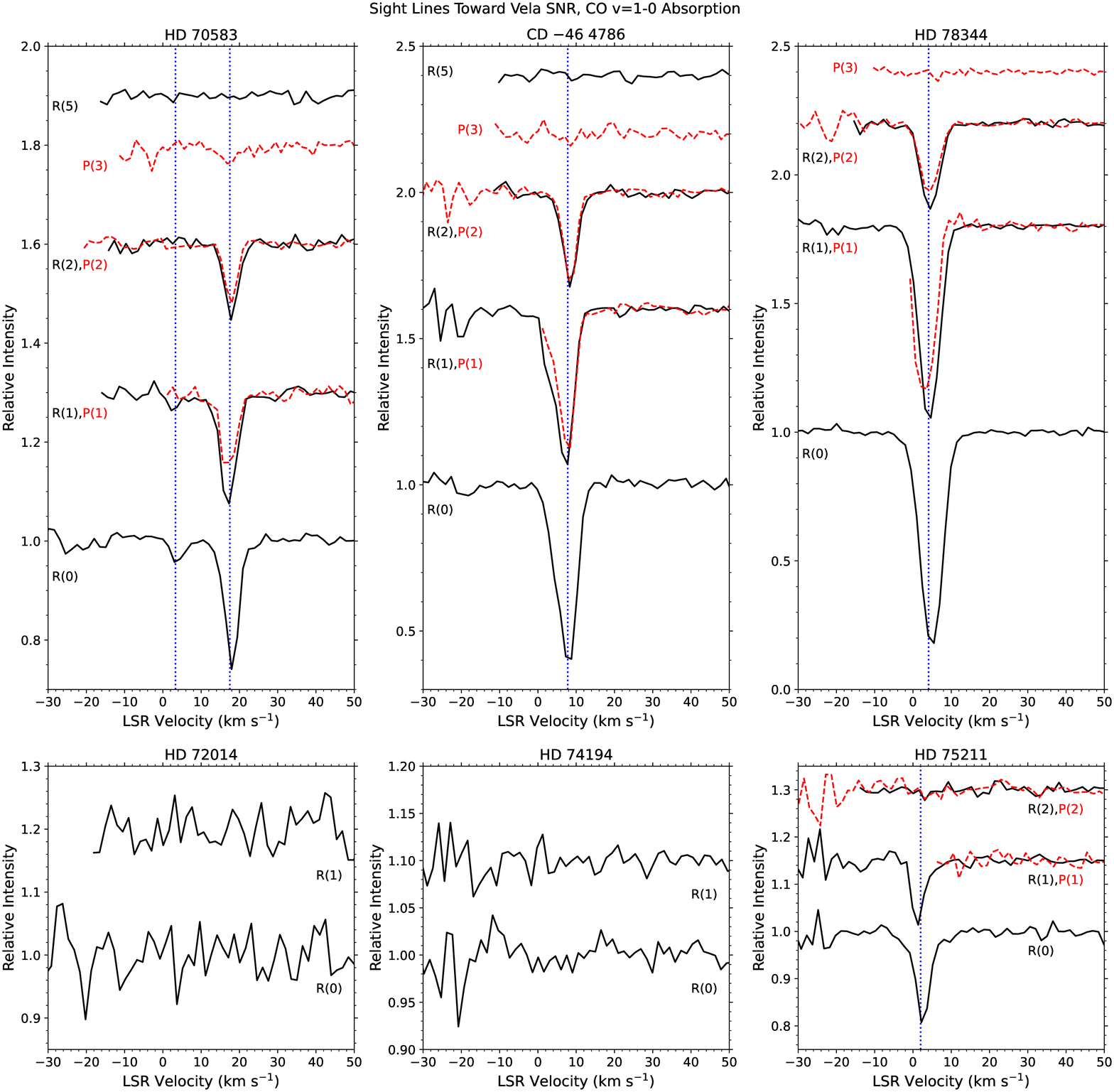}
\caption{Observations targeting the $v=1$--0 band of CO near 4.7~$\mu$m in six sight lines toward the Vela SNR---HD 70583, CD$-$46 4786, HD 78344, HD 72014, HD 74914, and HD 75211---are shown here. Notation is the same as in Figures \ref{fig_w28_spectra} and \ref{fig_vela_spectra}, but no H$_3^+$ observations were made toward these sight lines.}
\label{fig_vela_COspectra}
\end{figure}

To remove baseline fluctuations and atmospheric features the science target spectra were divided by the corresponding telluric standard star spectra using custom macros developed in IGOR Pro that allow for stretching and shifting of the telluric standard spectrum in the wavelength axis, and scaling of the telluric standard intensity according to Beer's law \citep{mccall2001}.  The resulting ratioed spectra were then divided by a 30 pixel boxcar average of the continuum level (interpolated across absorption lines) to remove residual fluctuations and produce normalized spectra. Regions where strong atmospheric absorption features result in no useful information are removed from the spectra to improve visualization. Because Earth's orbital motion causes astrophysical lines to shift with respect to atmospheric lines (i.e., observed wavelength) throughout the year, wavelength scales for all spectra were converted to the local standard of rest (LSR) frame. For any targets that were observed on multiple nights, the normalized spectra are combined using a weighted average (weighted by 1/$\sigma^2$, where $\sigma$ is the standard deviation of the line-free continuum in each spectrum). The normalized spectra resulting from these reduction procedures are presented in Figures \ref{fig_w28_spectra}--\ref{fig_vela_COspectra}. 

\section{Results and Analysis}

\subsection{H$_3^+$}
H$_3^+$ absorption is detected in 4 sight lines. Each absorption line is fit with a gaussian function to determine equivalent widths, line center velocities, and line widths. In the case of WR 105, each absorption feature is fit with the sum of two gaussian functions given the two velocity components. The resulting fit parameters are presented in Table \ref{tbl_measurements}, along with the column densities inferred from equivalent widths ($W_\lambda$) given the standard equation for optically thin absorption
\begin{equation}
N(J,K)=\left(\frac{3hc}{8\pi^3}\right)\frac{W_\lambda}{\lambda}\frac{1}{|\mu|^2},
\label{eq_column}
\end{equation}
where $N(J,K)$ is the column density in the state from which the transition arises, $h$ is Planck's constant, $c$ is the speed of light, $\lambda$ is the transition wavelength, and $|\mu|^2$ is the square of the transition dipole moment \citep[$\lambda$ and $|\mu|^2$ can be found in Table 2 of][]{mccall1999}.

\subsection{CO}
CO absorption is detected in 8 of the 11 sight lines presented in this study. As for H$_3^+$, each CO absorption line is fit with a gaussian function to determine equivalent widths, line center velocities, and line widths. Two components are required for WR 105, CD$-23$~13793, and HD 70583. Fit parameters are presented in Table \ref{tbl_measurementsCO}, as well as column densities for individual rotational states of CO computed in the optically thin limit as
\begin{equation}
N(J)_{thin}=8\pi c\frac{g_l}{g_u}\frac{W_{\lambda}}{A_{ul}\lambda^4},
\label{eq_column_CO}
\end{equation}
where $g_l$ and $g_u$ are the statistical weights of the lower and upper states, respectively, and $A_{ul}$ is the spontaneous emission coefficient of the transition. Transition data were taken from the HITRAN database \citep{hitran2012}. For CO transitions where no absorption is detected, upper limits to the equivalent width are computed as 
\begin{equation}
W_\lambda=\sigma\lambda_{\rm pix}\sqrt{{\cal N}_{\rm pix}},
\label{eq_equivwidul}
\end{equation}
where $\sigma$ is the standard deviation on the continuum level near the expected CO lines, $\lambda_{\rm pix}$ is the wavelength coverage per pixel, and ${\cal N}_{\rm pix}$ is the number of pixels expected in an absorption feature. Upper limits on the CO column densities in these rotational states are then calculated using equation (\ref{eq_column_CO}), and are also presented in Table \ref{tbl_measurementsCO}.

Column densities computed from transitions probing the same rotational state (e.g., $R(1)$ and $P(1)$) should produce consistent results. In several cases though---the $J=1$ and $J=2$ states in the 4~km~s$^{-1}$ component toward WR~105, the $J=1$ state toward HD~313599, the $J=1$ state in the 17~km~s$^{-1}$ component toward HD~70583, the $J=1$ and $J=2$ states toward CD $-$46 4786, and the $J=1$ and $J=2$ states toward HD~78344---the column densities determined from the $R$ and $P$ branch transitions do not agree within uncertainties. These discrepancies may indicate unaccounted for systematic uncertainties in the measured equivalent widths, or that the assumption of optically thin absorption is not valid. It is likely that both possibilities are contributing to varying degrees in each sight line, and I start by considering the latter.

To avoid the assumption of optically thin absorption, a curve-of-growth analysis is used to compute column densities from the measured equivalent widths. The column density in each rotational state is computed using the measured $W_{\lambda}$ for each transition as a function of the Doppler parameter, $b$, over the range $0.2\leq b\leq 2$~km~s$^{-1}$. For some value of $b$, $N(J)$ computed from both the $R(J)$ and $P(J)$ equivalent widths should be equal, and this represents the actual value of $N(J)$. The Doppler parameter is a property of the interstellar cloud being probed, so $b$ should be the same when derived from different rotational states (i.e., $b$ derived from the $R(1)$ and $P(1)$ lines should agree with that derived from the $R(2)$ and $P(2)$ lines), assuming molecules in both quantum states are distributed in the same way throughout the cloud. For WR~105, CD$-$46~4786, and HD~78344 the variance weighted averages of the Doppler parameters derived from the $J=1$ and $J=2$ transitions are adopted as the $b$-values for each sight line ($b=1.19\pm0.1$~km~s$^{-1}$, $b=1.26\pm0.2$~km~s$^{-1}$, and $b=1.16\pm0.2$~km~s$^{-1}$, respectively), and used in computing the column densities presented in the last two columns of Table \ref{tbl_measurementsCO}. Note that the value of $b=1.16\pm0.2$~km~s$^{-1}$ found for HD~78344 is in good agreement with the previous estimate of 1~km~s$^{-1}$ from an analysis of CO $J=1$--0 emission \citep{gredel2002}.

For HD~313599 and HD~70583 column densities derived from the $R(1)$ and $P(1)$ absorption lines do not agree within uncertainties, despite the fact that the absorption features are less than 20\% deep. For lines this shallow, the intrinsic line widths would have to be very narrow for absorption to no longer be in the optically thin regime. Indeed, the curve-of-growth analysis is incapable of finding a $b$-value for HD~313599 for which both $N(1)$ estimates agree, while HD~70583 requires $b=0.6\pm0.2$~km~s$^{-1}$. In both sight lines the $J=2$ column densities derived from $R(2)$ and $P(2)$ transitions are in agreement under the assumption of optically thin absorption, and that agreement gets worse if $b$ is set to improve the agreement in the $J=1$ column densities.

Considering potential systematic uncertainties in the measured equivalent widths, the most likely source comes from the process of dividing the science target spectrum by the telluric standard spectrum to remove atmospheric absorption features. In many cases the interstellar CO absorption features are located in the broad wings of strong atmospheric absorption lines, where it is difficult to determine the correct continuum level. This is particularly true for the $P(1)$ transition, as can be seen in Figures \ref{fig_w28_spectra}--\ref{fig_vela_COspectra} where the useable portion of the spectrum ends very near to the $P(1)$ absorption lines (i.e., the continuum level was poorly determined, resulting in a very noisy region of the spectrum that has been omitted here for ease of viewing the spectra). This means that the uncertainties on $W_{\lambda}$ quoted in Table \ref{tbl_measurementsCO}---returned by the guassian fitting process---are likely underestimates because they do not account for uncertainty in the continuum level chosen during the normalization process.

\subsection{Cosmic-Ray Ionization Rates \label{sect_crir}}

The simple chemistry by which H$_3^+$ is formed
\begin{equation}
{\rm H}_2 + {\rm CR}\rightarrow {\rm H}_2^+ + e^- + {\rm CR}',
\label{re_CR_H2}
\end{equation}
\begin{equation}
{\rm H}_2^+ + {\rm H}_2\rightarrow {\rm H}_3^+ + {\rm H},
\label{re_H2_H2+}
\end{equation}
and destroyed
\begin{equation}
{\rm H}_3^+ + e^-\rightarrow {\rm H}_2 + {\rm H~or~H + H + H},
\label{re_H3+_e}
\end{equation}
in diffuse clouds makes its abundance useful for inferring cosmic-ray ionization rates. While there are some minor reaction channels that make the picture more complex \citep[see, e.g.,][]{indriolo2012}, the analytical expression defined by assuming steady-state chemistry for reactions \ref{re_CR_H2}--\ref{re_H3+_e},
\begin{equation}
\zeta({\rm H}_2)=k_{e}x_{e}n_{\rm H}\frac{N({\rm H}_3^+)}{N({\rm H}_2)},
\label{eq_crir}
\end{equation}
is a reasonable approximation for the cosmic-ray ionization rate of H$_2$ in diffuse molecular clouds \citep{neufeld2017}. Here, $\zeta({\rm H}_2)$ is the total cosmic-ray ionization rate of H$_2$ (i.e., accounting for ionization by cosmic ray protons and secondary electrons), $k_e$ is the H$_3^+$-electron recombination rate coefficient, $n_{\rm H}\equiv n({\rm H})+2n({\rm H}_2)$ is the density of hydrogen nuclei, $x_{e}\equiv n_e/n_{\rm H}$ is the electron fraction, and $N({\rm H}_2)$ and $N({\rm H}_3^+)$ are column densities.

$N({\rm H}_2)$ has not been directly measured in any of the target sight lines, and so must be estimated. The preferred method for estimating $N({\rm H}_2)$ uses the measured value of $N({\rm CH})$ and the linear relationship between the abundances of both species \citep{sheffer2008}. If CH has not been observed, then $N({\rm H}_2)$ is estimated from the color excess, $E(B-V)$, and the relation $N_{\rm H}\approx E(B-V)5.8\times10^{21}$ cm$^{-2}$ mag$^{-1}$ from \citet{bohlin1978}, assuming equal amounts of H and H$_2$ along the sight line.
A complication arises with this method when there are multiple absorbing clouds along the line of sight, since color excess only provides information about the total amount of material in front of the background star. Considering each absorption component as an individual cloud with the chemical structure of a 1-D slab model illuminated on one side \citep[e.g.,][]{hollenbach1997,snow2006}, some reasonable limiting assumptions to make are (1) the component with lower $N({\rm CO})$ should have a CO/H$_2$ ratio that is smaller than the component with higher $N({\rm CO})$, and (2) the component with lower $N({\rm CO})$ should have less H$_2$ than the component with higher $N({\rm CO})$. These limiting cases (either both components have equal $N({\rm H}_2)$ or equal $N({\rm CO})/N({\rm H}_2)$) provide the limits on $N({\rm H}_2)$ shown in Table \ref{tbl_zeta} for WR 105 and HD 70583.

CO abundances can also be useful for constraining properties of the gas from which the absorption arises. In the 1-D chemical model of a cloud mentioned above, hydrogen transitions from H to H$_2$, while carbon transitions from C$^+$, to C, to CO going from the edge to the interior. These transitions mean that the relative abundance of CO with respect to total hydrogen changes throughout clouds, and that the relative abundances integrated through the cloud, i.e., $N({\rm CO})/N({\rm H}_2)$, will be different for clouds dominated by diffuse or dense gas. More diffuse clouds will have $N({\rm CO})/N({\rm H}_2)\sim10^{-6}$, while more dense clouds will have $N({\rm CO})/N({\rm H}_2)\sim10^{-4}$. As the H$_3^+$ chemistry described above is valid for diffuse molecular clouds \citep[100 cm$^{-3}\lesssim n_{\rm H} \lesssim$ 500 cm$^{-3}$, 30 K $\lesssim T\lesssim$ 100 K, $n({\rm C}^+) \geq (n({\rm C})+n({\rm CO}))$;][]{snow2006}, it is important to determine whether or not the clouds being studied fall into this regime. The $N({\rm CO})/N({\rm H}_2)$ ratios reported in Table \ref{tbl_zeta} indicate that most sight lines fall in the diffuse cloud regime, although a few are more likely probing translucent clouds (WR 105, CD $-46$ 4786, and HD 73882). The HD 78344 sight line appears to be probing a dense cloud. 

Hydrogen number density is a difficult parameter to determine for interstellar clouds. One method for constraining $n_{\rm H}$ utilizes the relative populations of different rotational states of a given molecule assuming that the level populations are predominantly controlled by collisional (de-)excitation with H and H$_2$ \citep[e.g., C$_2$;][]{sonnentrucker2007}. \citet{goldsmith2013} details an analysis of this sort using the $J=0$ through $J=3$ populations of CO. By calculating excitation temperatures from the state-specific CO column densities in Table \ref{tbl_measurementsCO} and utilizing expanded versions (Goldsmith 2015, private communication) of the $T_{ex}$ vs $n({\rm H}_2)$ curves in \citet{goldsmith2013} Figure 3, I infer $n_{\rm H}$ for the targeted sight lines. In several cases the densities estimated from $T_{10}$, $T_{21}$, and $T_{32}$ in a single sight line are not in agreement. This may be due changing conditions within the cloud such that the higher lying states are only populated in the cloud interior, while the lower lying states are populated everywhere. In these cases I take the mean value of $n_{\rm H}$ and adopt uncertainties equal to the difference between the largest density estimate and the mean. Gas densities estimated in this way are presented in Table \ref{tbl_zeta}. In the absence of CO observations, a value of 200~cm$^{-3}$ is adopted for $n_{\rm H}$ as was assumed in \citet{indriolo2012}. The density estimates for WR~104, WR~105, HD~70583, CD $-46$ 4786, and HD~78344 place these clouds in the translucent to dense regime, as suggested by the $N({\rm CO})/N({\rm H}_2)$ ratios (except for WR~104).

The electron recombination rate coefficient has been measured to be $k_e=-1.3\times10^{-8}+1.27\times10^{-6}T_e^{-0.48}$, where $T_e$ is the electron temperature \citep{mccall2004}. Assuming that nearly all electrons are the result of singly photoionized carbon, the electron fraction can be approximated by the fractional abundance of C$^+$, measured to be about $1.5\times10^{-4}$ in diffuse clouds \citep{cardelli1996,sofia2004}. However, this assumption can break down for two reasons discussed by \citet{neufeld2017}, each of which has an opposite effect on the electron fraction. First, as C$^+$ transitions to C from diffuse molecular to translucent clouds the electron abundance is no longer directly linked to the carbon abundance. In this case the average value of $x({\rm C}^+)$ will be an overestimate of $x_e$, leading to overestimates of $\zeta({\rm H}_2)$ as well. In some sight lines, $\zeta({\rm H}_2)$ inferred using the analytical methods described above may be 10 times higher than that inferred from more detailed chemical models \citep{neufeld2017}. Second, if the cosmic-ray ionization rate is high, then the abundance of electrons released by cosmic-ray ionization of H and H$_2$ can rival or surpass that produced by photoionization of C. In this case the average value of $x({\rm C}^+)$ will be an underestimate of $x_e$, leading to underestimates of $\zeta({\rm H}_2)$. Both of these effects are discussed in detail by \citet{neufeld2017} who find that the latter will dominate the former in setting $x_e$.

Given the preceding equations and assumptions, cosmic-ray ionization rates inferred from H$_3^+$ abundances are presented in the last column of Table \ref{tbl_zeta}. Of the resulting values, those for WR 104 and WR 105 are the most likely to be affected by $x_e$ overestimates as these sight lines pass through large columns of H$_2$ as estimated from $E(B-V)$, and may be probing denser gas. As such, the ionization rates inferred for WR~104 and WR~105 are not considered to be highly constraining in the discussion below. Properties of all other sight lines where H$_3^+$ is detected are consistent with the regime where cosmic-ray ionization rates inferred from the analytical expressions in \citet{indriolo2012} and detailed chemical models in \citet{neufeld2017} are in good agreement.

\section{Discussion}

\subsection{Vela SNR}

Unfortunately, H$_3^+$ observations toward 6 of the 8 target sight lines near the Vela SNR were never executed. To better study the cosmic-ray ionization rate in the region, I also consider previously reported H$_3^+$ observations toward HD~73882 \citep{crabtree2011,indriolo2012} as well as OH$^+$ observations toward HD 75309, HD 75860, and HD 79186 \citep{bacalla2019}, HD 78344 \citep{zhao2015}, and HD 76341 \citep{krelowski2010}. OH$^+$ is a tracer of the cosmic-ray ionization rate of atomic hydrogen in diffuse clouds, and the relationship between the primary ionization rate of H ($\zeta_p$; does not include ionization by secondary electrons), the total ionization rate of H, $\zeta({\rm H})$, and the total ionization rate of H$_2$ is roughly $\zeta_p=\zeta({\rm H})/1.5=\zeta({\rm H}_2)/2.3$ \citep{glassgold1973,glassgold1974}. Using this scaling relation, $\zeta({\rm H}_2)$ values corresponding to the $\zeta_p$ results reported in \citet{bacalla2019} are as follows: HD 75309: $(15.9\pm10.1)\times10^{-16}$~s$^{-1}$; HD 75860: $(11.5\pm1.6)\times10^{-16}$~s$^{-1}$; HD 78344:  $(15.2\pm1.6)\times10^{-16}$~s$^{-1}$; HD 79186: $(27.8\pm2.5)\times10^{-16}$~s$^{-1}$. Taking the values of $E(B-V)$ and $W_\lambda({\rm OH}^+)$ reported in \citet{krelowski2010} and using the analysis methods in \citet{bacalla2019}, the H$_2$ ionization rate toward HD 76341 is $(5.3\pm2.3)\times10^{-16}$~s$^{-1}$. Ionization rates inferred from OH$^+$ abundances are presented in Table \ref{tbl_onsky}. When combined with the results inferred from H$_3^+$, it is clear that all seven sight lines that pass through or near the Vela SNR show elevated cosmic-ray ionization rates with respect to the average value in diffuse clouds \citep[$\sim3\times10^{-16}$~s$^{-1}$;][]{indriolo2012,indriolo2015oxy}.

\begin{figure}
\epsscale{1.0}
\plotone{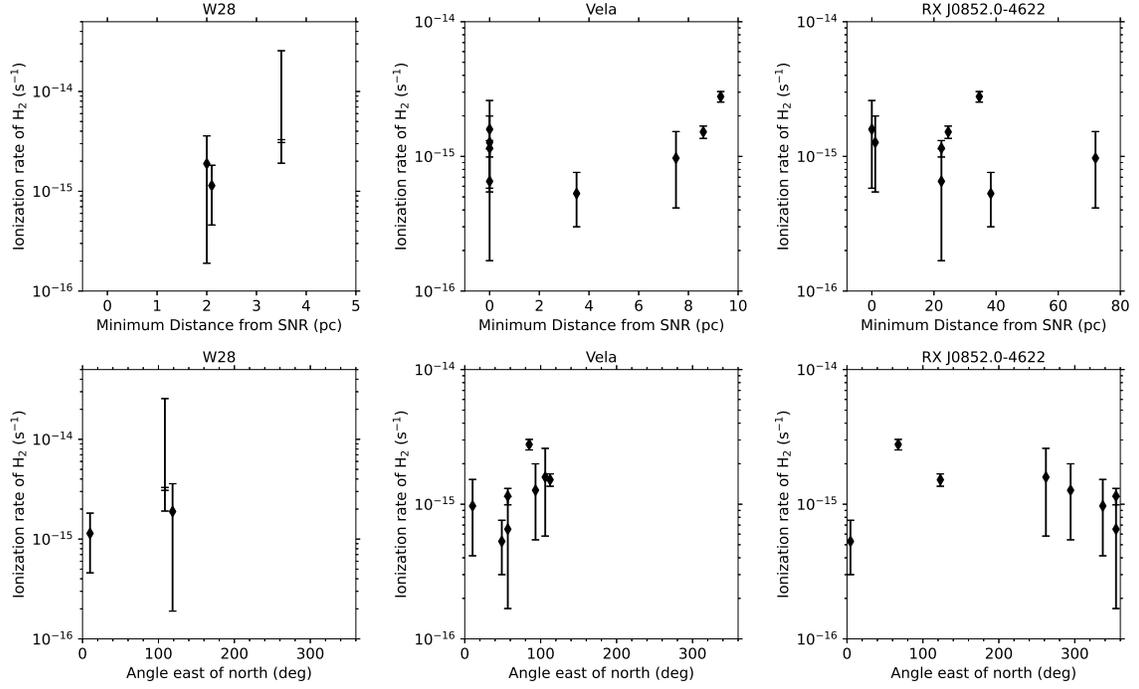}
\caption{In the top row, ionization rates are plotted as a function of the minimum distance between the absorbing gas and the SNR (values from Table \ref{tbl_onsky}). In the bottom row, ionization rates are plotted as a function of angle around the remnant, defined as $0^{\circ}$ at due north, and increasing to the east. Center coordinates used for each remnant to define these angles are 18h30m00s, $-23^{\circ}26'00''$ for W28, 08h34m00s, $-44^{\circ}50'00''$ for Vela, and 08h52m19.2s, $-46^{\circ}20'24''$ for RX J0852.0$-$4622.}
\label{fig_zeta_vs_dist}
\end{figure}

Figure \ref{fig_vela_map} shows the locations of the target sight lines with respect to the Vela SNR, RX J0852$-$4622, and Puppis A, while Table \ref{tbl_onsky} provides details about the on-sky distances between the sight lines and SNRs, as well as distances to the target stars themselves. There is no correlation between $\zeta({\rm H}_2)$ and on-sky separation from either Vela or RX J0852$-$4622, as can be seen in Figure \ref{fig_zeta_vs_dist}.
While the target stars are definitely beyond the Vela SNR, and most are also beyond the other two SNRs, the location of the absorbing gas along the line of sight is difficult to determine. This makes it challenging to definitively say whether or not the enhanced ionization rates are observed in gas that is in close proximity to SNRs. The best way to draw a connection between the absorbing gas and the gas near the SNRs is to search for similar radial velocities.

Absorption line studies of the Vela SNR targeting atomic gas (e.g., Na~\textsc{i} and Ca~\textsc{ii}) show several components over a wide range of velocities, some of which change in intensity or velocity over timescales of a few years \citep[e.g.,][]{cha2000,kameswara2020}. The high radial velocity gas ($\gtrsim50$~km~s$^{-1}$) is associated with the expanding Vela SNR, but these components tend to be much weaker than the deepest absorption at about 0--10~km~s$^{-1}$, which cannot be definitively linked to the SNR. Evidence for gas in the 10--30~km~s$^{-1}$ range interacting with the SNR comes in the form of high excitation C~\textsc{i} absorption probing several shocked dense clouds \citep{nichols2004}, but again this does not constrain the location of the 0--10~km~s$^{-1}$ gas where most of the CO and H$_3^+$ absorption arises. The best radial velocity match to the absorbing gas is CO emission from the VMR, which peaks in the 6--10~km~s$^{-1}$ range \citep{murphy1991,yamaguchi1999}. Distance estimates for the VMR are about 700$\pm$200~pc \citep{liseau1992}, much farther than the Vela SNR, and if the absorbing gas is associated with the Vela molecular clouds and their envelopes, then it would not be in close proximity to the SNR. Gaia distance estimates place HD~73882 at 358$_{-55}^{+79}$~pc \citep[DR2;][]{bailer-jones2018} or 755$_{-28}^{+28}$~pc \citep[EDR3;][]{bailer-jones2021}, with the latter potentially suffering from problems in the astrometric solution \citep{gaiaEDR32021}.  The sight line toward HD~73882 has an H$_2$ column density of $12.9\pm2.4\times10^{20}$~cm$^{-2}$ \citep{rachford2002}, and H$_3^+$ absorption centered at about 6~km~s$^{-1}$ \citep{indriolo2012}. If the smaller DR2 distance is adopted, then a substantial column of molecular gas must be in the foreground to the VMR and at the same radial velocity as the VMR, but if the DR3 distance is adopted, then this gas is likely part of the VMR. 
\citet{moriguchi2001} suggested that some of the CO emission coincident with the Vela SNR is truly in close proximity to the remnant due to the spatial anti-correlation of CO and X-ray emission in this region. Nothing conclusive about the location of absorbing gas along the target sight lines can be drawn from the radial velocities of H$_3^+$ and OH$^+$, but there is a possibility that the material is in close proximity to the Vela SNR.

Three sight lines considered here are in close proximity on-sky to the SNR RX J0852$-$4622: HD 75309, HD 75149, and CD $-$46 4786. RX J0852$-$4622 has a shell-type morphology observed in gamma rays \citep{aharonian2005,tanaka2011} that is well-matched by small clouds of CO and H~\textsc{i} emission in certain velocity intervals \citep{fukui2017}. This spatial correspondence between gamma-ray emission and interstellar matter is strong evidence for a hadronic rather than leptonic origin for the gamma rays. The gas velocity varies as a function of position in the remnant, and on the western edge CO emission is observed at about 24~km~s$^{-1}$. The HD 75309 sight line passes directly through the edge of the remnant in this region, while the HD 75149 sight line is nearby, just outside of the remnant. CO and H$_3^+$ absorption toward HD 75149 reported herein is centered at about 6~km~s$^{-1}$, so this gas is most likely associated with the VMR, not the small CO clouds.  OH$^+$ absorption toward HD 75309 is centered at 1.9~km~s$^{-1}$ (J. Cami, private communication) and so also not associated with the small clouds. CD $-46$ 4786 is located just outside of the southeast side of RX J0852$-$4622. Emission in the nearest small CO cloud peaks at about 20~km~s$^{-1}$, while the absorption is centered at about 8~km~s$^{-1}$, so again the absorbing material does not match the compact CO clouds. Despite the CO absorption velocities matching the VMR rather than the compact CO clouds interacting with RX J0852$-$4622, it is still possible that this material is in close proximity to the remnant since the SNR itself may be associated with the VMR \citep{slane2001,allen2015}.

Only two target sight lines, HD 70583 and HD 72014, are reasonably close to the Puppis A SNR on-sky, and of those two only HD 70583 is located behind the remnant. Puppis A shows gamma-ray emission on the northeast side of the SNR that is coincident with CO and H~\textsc{i} emission in the $v_{\rm LSR}=8$--20~km~s$^{-1}$ range \citep{hewitt2012,aruga2022}. This indicates the interaction between hadronic cosmic rays and the ambient ISM, and the total cosmic-ray energy estimated from the gamma-ray luminosity places the remnant in a phase where cosmic rays are escaping \citep{aruga2022}. While the sight line toward HD 70583 passes 11~arcmin (4.3~pc at a distance of 1.3~kpc) to the south of Puppis A, the strong CO absorption at 18~km~s$^{-1}$ is within the velocity range that suggests potential association with the SNR. Future observations of ionization rate tracers in this sight line could place further constraints on the cosmic-ray flux near Puppis A.

The discussion thus far has assumed that inferred ionization rates are due entirely to ionization by cosmic rays. As many supernova remnants emit X-rays, there is likely to be some contribution from X-ray ionization as well. \citet{yusef-zadeh2003} give a prescription for estimating the ionization rate due to X-rays right at the edge of an SNR. Taking the X-ray luminosities of the Vela SNR and Puppis A reported by \citet{silich2020} and the X-ray flux of RX J0852$-$4622 reported by \citet{takeda2016} and using the methods of \citet{yusef-zadeh2003} scaled up by a factor of 2 (for H$_2$ ionization), I compute the X-ray ionization rates at the edge of these SNRs. I find $3\times10^{-17}$~s$^{-1}$, $2\times10^{-18}$~s$^{-1}$, and $4\times10^{-15}$~s$^{-1}$ for Vela, RX J0852$-$4622, and Puppis A, respectively. These results suggest that the ionization rates presented in Table \ref{tbl_onsky} are primarily due to cosmic-rays, as the Vela and RX J0852$-$4622 X-ray ionization rates are factors of 10--1000 times smaller, and because sight lines with inferred ionization rates are several tens of pc away from Puppis A.

\subsection{W28 SNR}

As shown in Figure \ref{fig_w28_map}, the sight lines toward HD~313599, CD$-$23 13793, and WR~105 are in close proximity to the W28 SNR, as is the sight line toward WR~104 for which CO and H$_3^+$ absorption were presented by \citet{mccall2002}. The latter three sight lines also pass through the SNR G6.67-0.42. Gamma-ray emission (HESS J1801-233) is coincident with the bright 90~cm emission on the northeast side of of W28 where dense molecular gas is also located. CO emission extends southward from this point, across the overlap region between both SNRs, to the H~\textsc{ii} regions at $\delta=-24^{\circ}$ which are also gamma-ray sources \citep[HESS J1800-24;][]{aharonian2008,maxted2016}. While the Gaia EDR3 distance estimates for WR~104 and CD$-$23 13793 place both stars in front of W28, their renormalized unit weight error (RUWE) values are much larger than 1.4, indicating problems with the astrometric solutions \citep{gaiaEDR32021}. The CO absorption observed toward both stars suggests that they are in fact behind the molecular cloud on the eastern edge of the W28 SNR, and thus capable of probing material near the remnant. Table \ref{tbl_onsky} gives the on-sky separation between the target sight lines and the edge of the W28 SNR at a distance of 1.9~kpc. As for Vela, Figure \ref{fig_zeta_vs_dist}  shows no correlation between on-sky separation and $\zeta({\rm H}_2)$.

Ionization rates inferred for WR 104 and WR 105 are well above the average value in diffuse molecular clouds, although as mentioned above these may all be largely overestimated. The ionization rate toward HD 313599 is also well above average, and is the only sight line in the region for which $\zeta({\rm H}_2)$ is likely well constrained. CO and H$_3^+$ absorption velocities in all three sight lines \citep[WR 104 H$_3^+$ and CO at about 10~km~s$^{-1}$ and 21~km~s$^{-1}$, respectively; ][]{mccall2002} are in reasonable agreement with the range of CO emission observed in this region \citep[$-10$--25~km~s$^{-1}$;][]{nicholas2011}. Previous estimates of the cosmic-ray ionization rate in dense gas near W28 were made using observations of CO, HCO$^+$, and DCO$^+$ emission \citep{vaupre2014}. Results ranged from the average value in dense clouds to highly elevated (about $3\times10^{-17}$~s$^{-1}$ to $3\times10^{-15}$~s$^{-1}$), indicating varied ionization rates throughout the region. The H$_3^+$ results match the high side of this range, although it is possibile that the absorbing gas is not located near the SNR along the line of sight. G6.67-0.42 is estimated to be at a distance of 3.7~kpc \citep{wang2020}, but there is no information available regarding gas velocities associated with this SNR. It is thus difficult to draw any conclusions about whether or not H$_3^+$ absorption might arise from gas in close proximity to G6.67-0.42.

The ionization rate due to X-rays at the edge of W28 can also be calculated using the methods discussed above, and is in fact presented in \citet{yusef-zadeh2003} as 2.3$\times10^{-16}$~s$^{-1}$. However, there are some inconsistencies between this result and other entries in their Table 1, so I have recalculated the X-ray ionization rate starting from their X-ray luminosity. I find an H$_2$ ionization rate of 5$\times10^{-17}$~s$^{-1}$, well below the inferred ionization rates for sight lines in this region, suggesting that cosmic-rays are primarily responsible for the ionization of H$_2$ here.

\subsection{Spatial Variations in the Cosmic-Ray Flux}

The ionization rates discussed above (and presented in Tables \ref{tbl_zeta} and \ref{tbl_onsky}) lead to two questions: What causes the ionization rate in one region to be higher (or lower) than elsewhere? What causes variations in the ionization rate within a localized region?

The cosmic-ray spectrum at any point in the Galaxy is thought to be controlled by a few main factors. One common starting point is to assume a ``sea'' of Galactic cosmic rays---the result of particles accelerated by all sources within the Galaxy and their diffusion throughout the disk prior to escape---that sets the base particle spectrum. This spectrum can then be modified by, e.g., proximity to a site of cosmic-ray acceleration that boosts the local particle flux, propagation through and interaction with molecular material that causes energy losses \citep{padovani2009}, or magnetic effects that alter particle trajectories \citep[e.g.,][]{silsbee2018}. Recent work suggests that the cosmic-ray sea (at energies relevant for $\pi^0$ production) is relatively uniform throughout the Galaxy, and that regions experiencing higher particle fluxes are likely located close to acceleration sites \citep{aharonian2020,peron2021}. This conclusion fits very well with the interpretation that the elevated ionization rates found in the W28 and Vela regions are caused by proximity to particle accelerators, namely the supernova remnants discussed herein.

On smaller spatial scales (of order a few pc), there are several more effects that should be considered when investigating the cosmic-ray flux. These are the effects that can lead to variations of the cosmic-ray ionization rate within a given region. Simulations of particle acceleration predict higher efficiencies for quasi-parallel geometries (background magnetic field is nearly aligned with the shock normal) than for other orientations \citep{caprioli2014i}, suggesting that the particle flux escaping an SNR can vary with position around the remnant, as is seen for SN 1006 \citep[e.g.,][]{rothenflug2004}. Additionally, particles escaping a remnant are expected to preferentially diffuse along magnetic field lines \citep{nava2013}, so that we might expect sight lines along magnetic field lines that intersect an SNR to experience higher particle fluxes than sight lines for which the vector pointing to an SNR is perpendicular to magnetic field lines. Magnetic focusing and mirroring are two more effects related to magnetic fields that can alter the cosmic-ray flux \citep[e.g.,][]{kulsrud1969}. Focusing refers to the increased flux of particles that occurs as field lines---and the particles traveling along them---converge. Mirroring refers to the reflection of particles in this same converging field geometry as the increasing field strength causes pitch angles to increase and reverse the direction of particle propagation along the field lines. While both of these effects work simultaneously, mirroring dominates over focusing such that particle fluxes are reduced in regions of increased magnetic field strength \citep{padovani2011}. Models of particle propagation through a molecular cloud complex accounting for all of these effects found that the particle flux varies with the strength and orientation of the magnetic field, demonstrating that magnetic effects can cause variations in the ionization rate on small spatial scales \citep{owen2021}. 

Magnetic field orientations around the Vela and W28 SNRs are not particularly well-constrained, although both regions have been the subject of multiple polarization studies \citep{milne1974,dickel1976,milne1980,milne1987,duncan1996}. W28 shows field lines that run parallel to the bright radio arc on the north and east sides \citep{milne1987}, but the orientation at the target sight lines is unknown. Vela shows a twisted magnetic field structure, with polarization vectors rapidly changing direction across the remnant \citep{duncan1996}. In both cases, it is not clear how the field structures can be related to the inferred ionization rates. A simplified analysis considering the angle defined by the target star, the center of the SNR, and a point due north of the SNR center is shown in the bottom row of panels in Figure \ref{fig_zeta_vs_dist}. In the case of a uniform magnetic field, the maximum ionization rates might be expected in the directions parallel to the field lines, and minimum ionization rates would be offset from the two maxima by 90$^{\circ}$. Toward W28 and Vela the sight lines do not sample a wide enough range of angles to search for this pattern. In RX J0852$-$4622 however, there may be a hint of this effect as the ionization rates in sight lines near 90$^{\circ}$ and 270$^{\circ}$ are higher than the ionization rates in sight lines near 0$^{\circ}$. As the 180$^{\circ}$ region is not covered and the uncertainties on $\zeta({\rm H}_2)$ are rather large, more observations would be required to confirm this trend.

\section{Summary}

Observations targeting H$_3^+$ and CO absorption were made toward sight lines in close proximity on-sky to the W28 and Vela SNRs. CO/H$_2$ abundance ratios and hydrogen number densities inferred from CO excitation are indicative of diffuse molecular or translucent cloud conditions along most of the target sight lines, regimes where the cosmic-ray ionization rate can be inferred from H$_3^+$ abundances using simple analytical expressions. All three sight lines toward the W28 SNR where ionization rates are inferred have values that are above the average value in Galactic diffuse molecular clouds, $\zeta({\rm H}_2)\sim 3\times10^{-16}$~s$^{-1}$. These findings are consistent with a previous study of the region where elevated ionization rates were found at multiple locations within the dense gas \citep{vaupre2014}.
All seven of the sight lines toward the Vela SNR where ionization rates are inferred (from H$_3^+$ and OH$^+$) have values that range from about 2--10 times the average value. X-ray ionization rates inferred at the edges of the W28, Vela, and RX J0852$-$4622 supernova remnants are 10 to 1000 times lower than the ionization rates found in the target sight lines, suggesting that the ionization is dominated by cosmic rays. The results toward both SNRs support the scenario where cosmic rays are accelerated in the remnants before diffusing outward and interacting with the nearby interstellar medium, where the high flux of cosmic rays produces elevated ionization rates that are imprinted on the chemical abundances. A major caveat of this interpretation is the unknown location of the absorbing gas along each line of sight. While a given sight line may pass close to or even through a SNR on-sky, there is no guarantee that the gas from which ionization rates are calculated is in close proximity to the remnant. That said, the elevated ionization rates reported here certainly hint at some regional influence on the cosmic-ray flux, regardless of association with SNRs. Furthermore, the results suggest that ionization rates should be considered in the context of the environments surrounding the sight lines from which they were derived. Such an analysis might reveal the source of the large scatter in the distribution of inferred $\zeta({\rm H}_2)$ within the Galaxy.

NI thanks the anonymous referee for suggestions that significantly improved this paper, and acknowledges Adam Ritchey for calculating and providing the interstellar CH column densities observed toward HD 313599 and CD $-23$ 13793, Paul Goldsmith for providing model curves used in the $T({\rm CO})_{ex}$ vs $n({\rm H}_2)$ analysis, and Jan Cami and the EDIBLES team for providing spectra of HD 75309, HD 75860, and HD 79186 from which the radial velocity of OH$^+$ absorption was measured. Based on observations collected at the European Organisation for Astronomical Research in the Southern Hemisphere under ESO programmes 087.D-0066(A), 087.D-0066(B), 088.D-0074(A), and 088.D-0074(B).

\software{Astropy \citep{astropy2013}, CRIRES pipeline v2.3.3, IRAF \citep{tody1986,tody1993}, Matplotlib \citep{matplotlib2007}, Scipy \citep{scipy2019arxiv}, Swarp \citep{bertin2002}}

\bibliographystyle{aasjournal}
\bibliography{indy_master}

%%%%%%%%%%%%%%%%%%%%%%%%%%%%%%Tables%%%%%%%%%%%%%%%%%%%%%%%%%%%%%%%%%%%%%%%%%%%

\clearpage
\begin{deluxetable}{cccccccc}
%\rotate
\tabletypesize{\scriptsize}
\tablecaption{H$_3^+$ Absorption Line Parameters \label{tbl_measurements}}
\tablehead{\colhead{Target} & \colhead{Transition} & \colhead{$v_{\rm LSR}$} & \colhead{FWHM} & \colhead{$W_\lambda$} & \colhead{$\sigma(W_\lambda)$} & \colhead{$N(J,K)$} & \colhead{$\sigma(N(J,K))$} \\
 &  & \colhead{(km s$^{-1}$)} &  \colhead{(km s$^{-1}$)} & \colhead{($10^{-6} \mu$m)} & \colhead{($10^{-6} \mu$m)} & 
 \colhead{(10$^{13}$ cm$^{-2}$)} & \colhead{(10$^{13}$ cm$^{-2}$)}
}
\startdata
HD 313599 & $R$(1,1)$^u$ & 12.7 & 7.2 & 4.18 & 0.33 & 17.3 & 1.39 \\
HD 313599 & $R$(1,0)  & 12.3 & 7.2 & 5.17 & 0.33 & 13.1 & 0.84 \\
HD 313599 & $R$(1,1)$^l$ & 12.3 & 7.7 & 4.17 & 0.29 & 19.1 & 1.05 \\
WR 105 & $R$(1,1)$^u$ & 5.3 & 9.2 & 3.63 & 0.37 & 15.1 & 1.52 \\
WR 105 & $R$(1,1)$^u$ & 14.4 & 5.6 &1.36  & 0.30 & 5.64 & 1.23 \\
WR 105 & $R$(1,0)  & 5.1 & 6.5 & 3.08 & 0.29 & 7.79 & 0.74 \\
WR 105 & $R$(1,0)  & 14.1 & 7.8 & 2.71 & 0.31 & 6.86 & 0.79 \\
WR 105 & $R$(1,1)$^l$ & 6.1 & 8.4 & 2.88 & 0.24 & 13.2 & 1.10 \\
WR 105 & $R$(1,1)$^l$ & 15.7 & 6.8 & 1.89 & 0.23 & 8.67 & 1.04 \\
HD 75149 & $R$(1,1)$^u$ & 6.6 & 5.2 & 0.92 & 0.18 & 3.83 & 0.73 \\
HD 75149 & $R$(1,0) & 7.4 & 5.7 & 13.7 & 0.22 & 3.47 & 0.55 \\
HD 75149 & $R$(1,1)$^l$ & 6.2 & 5.9 & 1.00 & 0.14 & 4.59 & 0.63  \\
HD 75860 & $R$(1,1)$^u$ & 3.5 & 7.1 & 1.49 & 0.19 & 6.18 & 0.77 \\
HD 75860 & $R$(1,0)  & 2.7 & 7.4 & 1.81 & 0.19 & 4.58 & 0.48 \\
HD 75860 & $R$(1,1)$^l$ & 2.5 & 8.6 & 1.65 & 0.17 & 7.57 & 0.79 \\
\enddata
\tablecomments{Parameters are determined by fitting H$_3^+$ absorption lines with gaussian functions.}
\end{deluxetable}
\normalsize

\clearpage
\startlongtable
\begin{deluxetable}{cccccccccc}
%\rotate
\tabletypesize{\scriptsize}
\tablecaption{CO Absorption Line Parameters \label{tbl_measurementsCO}}
\tablehead{\colhead{Target} & \colhead{Transition} & \colhead{$v_{\rm LSR}$} & \colhead{FWHM} & \colhead{$W_\lambda$} & \colhead{$\sigma(W_\lambda)$} & \colhead{$N(J)_{\rm thin}$} & \colhead{$\sigma(N(J)_{\rm thin})$} & \colhead{$N(J)$} & \colhead{$\sigma(N(J))$} \\
 &  & \colhead{(km s$^{-1}$)} &  \colhead{(km s$^{-1}$)} & \colhead{($10^{-6} \mu$m)} & \colhead{($10^{-6} \mu$m)} & 
 \colhead{(10$^{14}$ cm$^{-2}$)} & \colhead{(10$^{14}$ cm$^{-2}$)} & \colhead{(10$^{14}$ cm$^{-2}$)} & \colhead{(10$^{14}$ cm$^{-2}$)}
}
\startdata
HD 313599       &$R(0)$     &12.0       &4.0                  &22.17      &2.36       &10.14      &1.08       & …         & …         \\
HD 313599       &$R(1)$     &10.9       &4.2                  &13.89      &1.64       &9.54       &1.12       & …         & …         \\
HD 313599       &$P(1)$     &10.7       &4.5                  &13.95      &2.38       &19.09      &3.26       & …         & …         \\
HD 313599       &$R(2)$     &12.2       &5.4                  &17.22      &2.79       &13.17      &2.13       & …         & …         \\
HD 313599       &$P(2)$     &11.0       &4.6                  &13.76      &2.25       &15.66      &2.56       & …         & …         \\
HD 313599       &$P(3)$     & …         &5.0\tablenotemark{a} & …         &2.53       & …         &2.69       & …         & …         \\
CD $-$23 13793  &$R(0)$     &5.2        &3.5                  &3.81       &0.98       &1.74       &0.45       & …         & …         \\
CD $-$23 13793  &$R(1)$     &4.5        &2.4                  &2.65       &0.94       &1.82       &0.65       & …         & …         \\
CD $-$23 13793  &$R(2)$     & …         &3.0\tablenotemark{a} & …         &1.05       & …         &0.80       & …         & …         \\
CD $-$23 13793  &$R(0)$     &18.9       &3.8                  &4.33       &1.03       &1.98       &0.47       & …         & …         \\
CD $-$23 13793  &$R(1)$     &18.9       &4.0                  &2.68       &1.19       &1.84       &0.82       & …         & …         \\
CD $-$23 13793  &$R(2)$     & …         &4.0\tablenotemark{a} & …         &1.21       & …         &0.93       & …         & …         \\
WR 105          &$R(0)$     &4.8        &5.9                  &58.40      &1.44       &26.71      &0.66       &118.00     &14.70      \\
WR 105          &$R(1)$     &4.6        &6.0                  &59.89      &1.08       &41.16      &0.74       &202.00     &19.10      \\
WR 105          &$P(1)$     &3.9        &5.5                  &51.64      &1.21       &70.66      &1.65       &213.00     &18.00      \\
WR 105          &$R(2)$     &4.8        &6.1                  &56.37      &1.29       &43.10      &0.99       &171.00     &18.00     \\
WR 105          &$P(2)$     &4.3        &5.5                  &49.61      &1.31       &56.46      &1.49       &155.00     &13.50      \\
WR 105          &$P(3)$     &4.4        &5.9                  &43.65      &1.01       &46.29      &1.07       &102.00     &5.95       \\
WR 105          &$R(4)$     &5.2        &6.7                  &20.38      &0.58       &16.88      &0.48       &22.10      &0.84       \\
WR 105          &$R(5)$     &6.3        &5.0                  &3.99       &0.47       &3.37       &0.40       &3.51       &0.44       \\
WR 105          &$P(5)$     &6.2        &3.4                  &3.97       &0.44       &3.96       &0.44       &4.13       &0.50       \\
WR 105          &$R(6)$     &7.2        &2.6                  &1.49       &0.48       &1.28       &0.41       &1.30       &0.43       \\
WR 105          &$P(6)$     & …         &3.0\tablenotemark{a} & …         &0.37       & …         &0.36       & …         & …         \\
WR 105          &$R(0)$     &16.5       &4.3                  &15.55      &1.21       &7.11       &0.55       & …         & …         \\
WR 105          &$R(1)$     &16.3       &4.2                  &19.30      &0.89       &13.26      &0.61       & …         & …         \\
WR 105          &$P(1)$     &15.9       &3.8                  &10.55      &0.99       &14.44      &1.35       & …         & …         \\
WR 105          &$R(2)$     &16.4       &4.7                  &14.75      &1.11       &11.28      &0.85       & …         & …         \\
WR 105          &$P(2)$     &16.0       &4.0                  &8.48       &1.09       &9.65       &1.24       & …         & …         \\
WR 105          &$P(3)$     &16.4       &4.9                  &5.42       &0.92       &5.75       &0.97       & …         & …         \\
WR 105          &$R(4)$     &15.6       &2.9                  &1.57       &0.38       &1.30       &0.31       & …         & …         \\
WR 105          &$R(5)$     & …         &4.0\tablenotemark{a} & …         &0.62       & …         &0.52       & …         & …         \\
HD 70583        &$R(0)$     &3.9        &3.9                  &2.96       &0.60       &1.35       &0.28       & …         & …         \\
HD 70583        &$R(1)$     &2.9        &3.5                  &2.03       &0.48       &1.39       &0.33       & …         & …         \\
HD 70583        &$R(2)$     & …         &4.0\tablenotemark{a} & …         &0.59       & …         &0.45       & …         & …         \\
HD 70583        &$R(0)$     &18.0       &4.3                  &18.16      &0.64       &8.31       &0.29       & …         & …         \\
HD 70583        &$R(1)$     &17.1       &4.9                  &18.44      &0.58       &12.67      &0.40       & …         & …         \\
HD 70583        &$P(1)$     &17.2       &4.8                  &12.21      &0.70       &16.71      &0.96       & …         & …         \\
HD 70583        &$R(2)$     &17.9       &4.2                  &10.39      &0.57       &7.95       &0.44       & …         & …         \\
HD 70583        &$P(2)$     &17.8       &3.6                  &7.24       &0.33       &8.24       &0.37       & …         & …         \\
HD 70583        &$P(3)$     &17.2       &5.5                  &2.88       &0.81       &3.06       &0.86       & …         & …         \\
HD 70583        &$R(5)$     & …         &5.0\tablenotemark{a} & …         &0.50       & …         &0.42       & …         & …         \\
HD 72014        &$R(0)$     & …         &5.0\tablenotemark{a} & …         &2.15       & …         &0.98       & …         & …         \\
HD 72014        &$R(1)$     & …         &5.0\tablenotemark{a} & …         &1.90       & …         &1.31       & …         & …         \\
HD 74194        &$R(0)$     & …         &5.0\tablenotemark{a} & …         &0.88       & …         &0.40       & …         & …         \\
HD 74194        &$R(1)$     & …         &5.0\tablenotemark{a} & …         &0.68       & …         &0.47       & …         & …         \\
HD 75149        &$R(0)$     &6.0        &11.0                 &7.83       &1.01       &3.58       &0.46       & …         & …         \\
HD 75149        &$R(1)$     &3.9        &7.3                  &2.40       &0.46       &1.65       &0.32       & …         & …         \\
HD 75149        &$R(2)$     & …         &9.0\tablenotemark{a} & …         &0.43       & …         &0.33       & …         & …         \\
HD 75211        &$R(0)$     &2.6        &4.5                  &13.84      &1.05       &6.33       &0.48       & …         & …         \\
HD 75211        &$R(1)$     &1.3        &3.6                  &8.02       &0.85       &5.51       &0.58       & …         & …         \\
HD 75211        &$R(2)$     & …         &4.0\tablenotemark{a} & …         &0.59       & …         &0.45       & …         & …         \\
HD 75860        &$R(0)$     & …         &7.0\tablenotemark{a} & …         &0.45       & …         &0.21       & …         & …         \\
HD 75860        &$R(1)$     & …         &7.0\tablenotemark{a} & …         &0.64       & …         &0.44       & …         & …         \\
CD $-$46 4786   &$R(0)$     &7.5        &6.3                  &63.59      &2.42       &29.08      &1.11       &143.00     &31.10      \\
CD $-$46 4786   &$R(1)$     &6.6        &6.4                  &54.19      &2.12       &37.24      &1.46       &112.00     &16.40      \\
CD $-$46 4786   &$P(1)$     &7.4        &5.4                  &41.86      &1.42       &57.28      &1.95       &111.00     &8.58       \\
CD $-$46 4786   &$R(2)$     &8.4        &4.3                  &22.31      &0.73       &17.06      &0.56       &22.50      &1.00       \\
CD $-$46 4786   &$P(2)$     &8.4        &3.9                  &20.33      &0.64       &23.13      &0.73       &29.50      &1.23       \\
CD $-$46 4786   &$P(3)$     &8.1        &4.0                  &2.35       &1.19       &2.49       &1.26       & …         & …         \\
CD $-$46 4786   &$R(5)$     & …         &4.0\tablenotemark{a} & …         &0.67       & …         &0.57       & …         & …         \\
HD 78344        &$R(0)$     &4.9        &6.7                  &93.31      &0.98       &42.68      &0.45       &7060.00    &1070.00    \\
HD 78344        &$R(1)$     &4.3        &6.1                  &77.46      &0.72       &53.23      &0.50       &1380.00    &123.00     \\
HD 78344        &$P(1)$     &3.0        &6.1                  &70.11      &1.75       &95.92      &2.40       &1160.00    &237.00     \\
HD 78344        &$R(2)$     &4.5        &5.6                  &31.14      &0.91       &23.81      &0.70       &38.50      &1.95       \\
HD 78344        &$P(2)$     &4.3        &5.6                  &25.66      &1.22       &29.20      &1.39       &42.00      &3.29       \\
HD 78344        &$P(3)$     & …         &6.0\tablenotemark{a} & …         &0.94       & …         &1.00       & …         & …         \\
\enddata
\tablecomments{Parameters determined from fitting CO absorption lines with gaussian functions. In the case of non-detections, the 1$\sigma$ limits are reported. For WR 105, CD $-$46 4786, and HD 78344 doppler parameters of $b=1.19$~km~s$^{-1}$, $b=1.26$~km~s$^{-1}$, and $b=1.16$~km~s$^{-1}$, respectively, were adopted when computing column densities using a curve-of-growth analysis.}
\tablenotetext{a}{FWHM adopted for computing upper limits on the equivalent width in the case of non-detections.}
\end{deluxetable}
\normalsize

\clearpage
\begin{deluxetable}{cccccccc}
%\rotate
\tabletypesize{\scriptsize}
\tablecaption{Total Column Densities and Cosmic-ray Ionization Rates \label{tbl_zeta}}
\tablehead{\colhead{Target} & \colhead{$N({\rm H}_2)$} &  \colhead{H$_2$ Reference} & \colhead{$N({\rm CO})$} & \colhead{$N({\rm CO})/N({\rm H}_2)$} & \colhead{$n_{\rm H}$} & \colhead{$N({\rm H_3^+})$} &  \colhead{$\zeta({\rm H}_2)$} \\
 & \colhead{(10$^{20}$ cm$^{-2}$)} &   & \colhead{(10$^{14}$ cm$^{-2}$)} & \colhead{($\times10^{-6}$)} & \colhead{(cm$^{-3}$)} & \colhead{(10$^{13}$ cm$^{-2}$)} &  \colhead{($10^{-16}$ s$^{-1}$)} 
}
\startdata
\multicolumn{8}{l}{W28 SNR} \\
HD 313599 & 10.1$\pm$5.06 & CH,1 & 34.9$\pm$2.2 & 3.4 & 160$\pm$40 & 31.6$\pm$1.52 & 11.4$\pm$6.81 \\
CD $-$23 13793 (5~km~s$^{-1}$) & 11.4$\pm$5.6 & CH,1 & 3.56$\pm$0.79 & 0.31 & 160$\pm$110 & ... & ... \\
CD $-$23 13793 (19~km~s$^{-1}$) &  5.6$\pm$2.8 & CH,1 & 3.82$\pm$0.94 & 0.68 & 190$\pm$120 & ... & ... \\
WR 105 (5~km~s$^{-1}$)  & 23.3$\leq N({\rm H}_2)\leq$43.9  & $E(B-V)$,2  & 616$\pm$23 & 14--26 & 1600$\pm$850 & 21.6$\pm$1.50 & 18.1--33.9\tablenotemark{a}   \\
WR 105 (15~km~s$^{-1}$) & 2.73$\leq N({\rm H}_2)\leq$23.3 & $E(B-V)$,2 & 38.4$\pm$1.5 & 1.7--14 & 2220$\pm$1540 & 14.3$\pm$2.28 & 30.8--256\tablenotemark{a}   \\
WR 104 & 44.7$\pm$22.3 & $E(B-V)$,3 & 94.9$\pm$2.2 & 2.1 & 1590$\pm$1130 & 23.2$\pm$1.5 & 18.9$\pm$17.0\tablenotemark{a}  \\
\hline
\multicolumn{8}{l}{Vela SNR} \\
HD 70583 (3~km~s$^{-1}$) & 0.99$\leq N({\rm H}_2)\leq$6.38 & $E(B-V)$,4 & 2.75$\pm$0.43 & 0.43--2.8 & 160$\pm$80 & ... & ... \\
HD 70583 (17~km~s$^{-1}$) & 6.38$\leq N({\rm H}_2)\leq$11.8 & $E(B-V)$,4 & 32.5$\pm$1.0 & 2.8--5.1 & 3750$\pm$2550 & ... & ... \\
HD 72014 & 2.90$\pm$1.45 & $E(B-V)$,5 & $<6.87$ & $<2.4$ & ... & ... & ... \\
HD 74194 & 3.71$\pm$1.83 & CH,6 &  $<2.61$ & $<0.70$ & ... & ... & ... \\
HD 75149 & 4.86$\pm$2.43 & CH,6  & 5.23$\pm$0.56 & 1.1 & 350$\pm$50 & 7.73$\pm$0.77 & 12.7$\pm$7.26 \\
HD 75211 & 6.57$\pm$3.24 & CH,6 & 11.8$\pm$0.8 & 1.8 & 210$\pm$40 & ... & ... \\
HD 75860 & 8.00$\pm$3.98 & CH,6  & $<1.93$ & $<0.24$ & 200 & 11.4$\pm$1.09 & 6.53$\pm$4.85 \\
CD $-$46 4786 & 17.1$\pm$8.80 & CH,6 & 280$\pm$32 & 16 & 1730$\pm$1480 & ... & ... \\
HD 78344  & 17.1$\pm$8.45 & CH,6 & 8430$\pm$1080 & 492 & 3090$\pm$2600 & ... & ... \\
HD 73882 & 12.9$\pm$2.39 & H$_2$,7 & 355$\pm$170 & 28 & 520 & 9.02$\pm$0.50 & 9.71$\pm$5.57 \\
\enddata
\tablecomments{H$_2$ column densities are estimated using proxy observations as indicated in the H$_2$ References column. An entry of `CH' signifies that the H$_2$ column density is calculated from the CH column density using the relation $N({\rm CH})/N({\rm H}_2)=3.5^{+2.1}_{-1.4}\times10^{-8}$ from \citet{sheffer2008}. An entry of `$E(B-V)$' signifies that the H$_2$ column density is calculated from $E(B-V)$ using the relation $N_{\rm H}\approx E(B-V)5.8\times10^{21}$ cm$^{-2}$ mag$^{-1}$ from \citet{bohlin1978}, and assuming $2N({\rm H}_2)/N_{\rm H}=f_{{\rm H}_2}=2/3$. An entry of `H$_2$' indicates direct measurement from H$_2$ absorption. Numbers refer to the sources of $N({\rm CH})$ and $E(B-V)$, and are as follows: (1) Adam Ritchey, private communication; (2) - \citet{rate2020}; (3) - \citet{conti1990}; (4) - $E(B-V)$ estimated from photometry \citep{zacharias2012}, spectral type \citep{houk1978,houk1993}, and intrinsic color \citep{fitzgerald1970}; (5) - \citet{jenkins1976}; (6) - \citet{gredel2002}; (7) - \citet{rachford2002}. Total CO and H$_3^+$ column densities are found by summing over the state-specific column densities reported in Tables \ref{tbl_measurements} and \ref{tbl_measurementsCO}. In cases where a state is probed by multiple transitions (e.g., the $J=1$ and $J=2$ states of CO, and the $(J,K)=(1,1)$ state of H$_3^+$) the state-specific column density is determined from a variance weighted average of the different measurements before being combined in the sum over all states. In cases where CO absorption is not detected, 3$\sigma$ upper limits on column densities are reported. CO and H$_3^+$ column densities for WR 104 are from \citet{mccall2002}. For HD 73882 $N({\rm CO})$ and $n_{\rm H}$ are from \citet{sonnentrucker2007} and $N({\rm H}_3^+)$ is from \citet{crabtree2011}.
}
\tablenotetext{a}{This ionization rate may be an overestimate as diffuse cloud chemistry becomes a poor assumption at the inferred gas density}
\end{deluxetable}
\normalsize

\clearpage
\begin{deluxetable}{cccccccccc}
\rotate
\tabletypesize{\scriptsize}
\tablecaption{On-sky separation between target sight lines and SNRs \label{tbl_onsky}}
\tablehead{\colhead{Target} & \colhead{$\zeta({\rm H}_2)$} & \colhead{$d$ Gaia DR2} & \colhead{$d$ Gaia EDR3} & \colhead{$v_{\rm LSR}({\rm H}_3^+)$} & \colhead{$v_{\rm LSR}({\rm OH}^+)$} & \colhead{Angular Separation} & \colhead{On-Sky Separation} & \colhead{Angular Separation} & \colhead{On-Sky Separation}  \\
  &  \colhead{($10^{-16}$ s$^{-1}$)} & \colhead{(pc)} & \colhead{(pc)} & \colhead{(km s$^{-1}$)} & \colhead{(km s$^{-1}$)} & (arcmin) & (pc) & (arcmin) & (pc) }
\startdata
 & & & & & & \multicolumn{2}{c}{W28 SNR} & \multicolumn{2}{c}{G6.67$-$0.42} \\
 \hline
HD 313599 & 11.4$\pm$6.81 & 2170$_{-258}^{+336}$ & 2230$_{-150}^{+193}\dagger$ & 12.4 & ... & 3.8 & 2.1 & N/A & N/A \\
CD $-$23 13793 & ... & 1052$_{-484}^{+2957}$ & 426$_{-45}^{+85}\dagger$ & ... & ... & 3.5 & 2.0 & T & 0 \\
WR 105 & 18.1--33.9, 30.8--256 & 1760$_{-205}^{+266}$ & 4131$_{-461}^{+556}$ & 5.5, 14.7 & ... & 6.3 & 3.5 & T & 0 \\
WR 104 & 18.9$\pm$17.0 & 3642$_{-1023}^{+1922}$ & 689$_{-99}^{+106}\dagger$ & 10.0 & ... & 3.7 & 2.0  & T & 0 \\
\hline
 & & & & & & \multicolumn{2}{c}{Vela SNR} & \multicolumn{2}{c}{RX J0852.0$-$4622}  \\
 \hline
HD 70583 & ... & 2023$_{-136}^{+156}$ & 2143$_{-66}^{+83}$ & ... & ... & T & 0 & 315 & 69 \\
HD 72014 & ... & 783$_{-39}^{+44}$ & 717$_{-22}^{+20}$ & ... & ... & T & 0 & 282 & 62  \\
HD 74194 & ... & 2212$_{-141}^{+160}$ & 2203$_{-88}^{+78}$ & ... & ... & T & 0 & 88 & 19.2  \\
HD 75149 & 12.7$\pm$7.26 & 1705$_{-229}^{+309}$ & 1466$_{-123}^{+122}$ & 6.7 & ... & T & 0 & 5 & 1.1  \\
HD 75211 & ... & 1716$_{-108}^{+123}$ & 1542$_{-36}^{+34}$ & ... & ... & T & 0 & 94 & 20.5  \\
HD 75860 & 6.53$\pm$4.85, 11.5$\pm$1.6 & 2035$_{-155}^{+182}$ & 2265$_{-195}^{+165}\dagger$ & 2.9 & 4.3 & T & 0 & 103 & 22.4  \\
CD $-$46 4786 & ... & 2121$_{-142}^{+163}$ & 2149$_{-47}^{+75}$ & ... & ... & 5 & 0.4 & 13 & 2.8  \\
HD 78344 & 15.2$\pm$1.6 & 2509$_{-232}^{+283}$ & 2149$_{-71}^{+83}$ & ... & N.R. & 103 & 8.6 & 113 & 24.6  \\
HD 73882 & 9.71$\pm$5.57 & 358$_{-55}^{+79}$ & 755$_{-28}^{+28}\dagger$ & 5.7 & ... & 90 & 7.5 & 327 & 72  \\
HD 75309 & 15.9$\pm$10.1 & 1931$_{-146}^{+171}$ & 1797$_{-105}^{+124}$ & ... & 1.9 & T & 0 & T & 0  \\
HD 76341 & 5.3$\pm$2.3 & 1292$_{-120}^{+146}$ & 1125$_{-65}^{+68}$ & ... & N.R. & 41 & 3.5 & 175 & 38.3 \\
HD 79186 & 27.8$\pm$2.5 & 1279$_{-238}^{+369}$ & 1677$_{-201}^{+212}\dagger$ & ... & 3.1 & 111 & 9.3 & 159 & 34.6  \\
\enddata
\tablecomments{In considering whether or not sight lines can probe material in close proximity to each SNR, the following distances are adopted: W28 SNR - 1.9~kpc; G6.67$-$0.42 - 3.7~kpc; Vela SNR - 287~pc; RX J0852.0$-$4622 - 750~pc. Columns 3 and 4 give the distances to the background stars as determined from the Gaia DR2 \citep{bailer-jones2018,gaiaDR22018} and EDR3 \citep{bailer-jones2021,gaiaEDR32021} catalogs, respectively. EDR3 entries for which the renormalized unit weight error (RUWE) is greater than 1.4 are marked with a $\dagger$ in column 4, as this indicates potential problems with the astrometric solution \citep{gaiaEDR32021}, and thus unreliable distance estimates. If a star is closer to the Earth than an SNR, an entry of N/A indicates that the sight line cannot be probing material in close proximity to that SNR. If a star is farther away from the Earth than an SNR and is located within the SNR on-sky, then an entry of T indicates that the sight line passes {\it through} the SNR. If a star is farther away from the Earth than an SNR and is located outside of the SNR on-sky, then the angular separation between the target star and the nearest edge of the SNR is reported. This angular separation is used along with the distance to the SNR to calculate the minimum possible distance between the absorbing gas and the SNR. For sight lines where H$_3^+$ or OH$^+$ absorption is detected, the mean LSR velocity of all absorption lines is reported. An entry of N.R. indicates that the gas velocity was not reported in the publication despite absorption being detected, while ... indicates that the molecule has not been observed. The two ionization rates reported for HD 75860 were calculated from H$_3^+$ and OH$^+$ abundances, respectively.}
\end{deluxetable}
\normalsize

%%%%%%%%%%%%%%%%%%%%%%%%%%%%%%Figures%%%%%%%%%%%%%%%%%%%%%%%%%%%%%%%%%%%%%%%%%%

\clearpage

\clearpage

\clearpage

\clearpage

\clearpage

\clearpage

\end{document}